\documentclass{PoS}

\usepackage{amsfonts}
\usepackage{amsmath}
\usepackage{subfigure} % Include figure files

\graphicspath{{./figs/}}
\title{ 
  Trouble shooting for covariance fitting in highly correlated data
}

\ShortTitle{
  Highly correlated data fitting 
}

\author{\speaker{Boram Yoon}, Yong-Chull Jang, Weonjong Lee\\
  Lattice Gauge Theory Research Center, CTP, and FPRD, \\
  Department of Physics and Astronomy,
  Seoul National University, Seoul, 151-747, South Korea \\
  E-mail: \email{wlee@snu.ac.kr}}

\author{Chulwoo Jung \\
  Physics Department, Brookhaven National Laboratory,
  Upton, NY11973, USA \\
  E-mail: \email{chulwoo@bnl.gov}}

\author{SWME Collaboration}

\abstract{
  We report a possible solution to the trouble that the covariance
  fitting fails when the data is highly correlated and the covariance
  matrix has small eigenvalues.
  As an example, we choose the data analysis of highly correlated
  $B_K$ data on the basis of the SU(2) staggered chiral perturbation
  theory.
  Basically, the essence of the problem is that we do not have an
  accurate fitting function so that we cannot fit the highly
  correlated and precise data.
  When some eigenvalues of the covariance matrix are small, even a
  tiny error of fitting function can produce large chi-square and
  spoil the fitting procedure.
  We have applied a number of prescriptions available in the market
  such as diagonal approximation and cutoff method.
  In addition, we present a new method, the eigenmode shift method
  which fine-tunes the fitting function while keeping the covariance
  matrix untouched.  }

\FullConference{
The XXIX International Symposium on Lattice Field Theory - Lattice 2011\\
July 10-16, 2011\\
Squaw Valley, Lake Tahoe, California
}

\begin{document}

%%%%%%%%%%%%%%%%%%%%%%%%%%%%%%%%%%
%%% SECTION :  Introduction    %%%
%%%%%%%%%%%%%%%%%%%%%%%%%%%%%%%%%%
\section{Introduction} 
We have reported results of $B_K$ calculated using improved staggered
fermions with $N_f=2+1$ flavors in Ref.~\cite{wlee-10-3}.
In Ref.~\cite{wlee-10-3}, we use the diagonal approximation
(uncorrelated fitting) instead of the full covariance fitting.
This is due to the fact that the $\chi^2$ value was out of range,
which indicates that the full covariance fitting fails manifestly.
One of the most frequently asked questions on Ref.~\cite{wlee-10-3} is
why we do the uncorrelated fitting instead of the full covariance fitting.

Here, we provide an elaborate answer to why we use the diagonal
approximation.
In addition, we propose a new method, named the eigenmode shift (ES)
method, which fine-tunes the fitting function while keeping the 
covariance matrix untouched.
More details on this issue will be reported in
Ref.~\cite{wlee-2011-1}.

%%%%%%%%%%%%%%%%%%%%%%%%%%%%%%%%%%%%%%
%%% SECTION :  Covariance fitting  %%%
%%%%%%%%%%%%%%%%%%%%%%%%%%%%%%%%%%%%%%
\section{Covariance fitting}
First, we review the covariance fitting. Then, we would like to address
the possible failure of the covariance fitting, which originates from 
the truncation error of the fitting function in the series expansion
of the staggered chiral perturbation theory (SChPT).

Let us consider $N$ samples of unbiased estimates of quantity 
$y_i$ with $i=1,2,3,\ldots,D$.
Here, the data set is $\{y_i(n) | n=1,2,3,\ldots,N \}$.
Let us assume that the samples $y_i(n)$ are statistically independent
in $n$ for fixed $i$ but are substantially correlated in $i$.
An introduction to this subject is given in
Ref.~\cite{milc-1988-1,toussaint-1990-1,anderson-2003,johnson-2007}.

We are interested in the probability distribution of the average
$\bar{y}_i$ of the data $y_i(n)$, defined by 
$\bar{y}_i = \frac{1}{N} \sum_{n=1}^{N} y_i(n)$.
We assume that the measured values of $\bar{y}_i$ have a normal
distribution $P(\bar{y})$ by the central limit theorem for the
multivariate statistical analysis as follows:
\begin{equation}
\label{eq:pdf_normal}
P(\bar{y}) = \frac{1}{Z} \exp\left[ -\frac{1}{2} \sum_{i,j=1}^{D}
(\bar{y}_i - \mu_i) (N \ \Gamma^{-1}_{ij}) (\bar{y}_j - \mu_j) 
\right] \,,
\end{equation}
where $\mu_i$ represents the true mean value of $y_i$, which is, in
general, unknown and can be obtained as $N \rightarrow \infty$, 
and $Z$ is the normalization constant.
Here, $\Gamma_{ij}$ is the true covariance matrix, which is, in
general, unknown in our problems.
The maximum likelihood estimator of $\dfrac{1}{N}\Gamma_{ij}$ turns out
to be the sample covariance matrix of mean, $C_{ij}$, defined as
follows,
\begin{eqnarray}
C_{ij} &=& \frac{1}{N(N-1)} \sum_{n=1}^{N} [y_i(n) - \bar{y}_i]
  [y_j(n) - \bar{y}_j]
\,.
\label{eq:cov_mat}
\end{eqnarray}

Let us consider a fitting function, $f_\text{th}(X_i; c_a)$. Here,
$X_i$ are the input variables which define data points and $c_a$ are
fitting parameters.  What we want to do is to determine the fitting
parameters to give the best fit and to test whether the fitting
function describes the data reliably from the standpoint of
statistics.
Here, the best fit is defined by minimizing the $T^2$, where $T^2$ is
\begin{eqnarray}
\label{eq:def_Tsq}
T^2 = \sum_{i,j=1}^{D}
[\bar{y}_i - f_\text{th}(X_i)] [C^{-1}_{ij}] [\bar{y}_j - f_\text{th} (X_j)]
\,.
\end{eqnarray}
In ideal case, the best fit gives the true mean of the 
data, $\mu_i$, in Eq.~\eqref{eq:pdf_normal}.
We notice that $\sqrt{N} [\bar{y}_i - f_\text{th} (X_i)]$ is distributed
according to the multivariate normal distribution, 
$\mathcal{N}(\rho, \Gamma)$, where 
$\rho_i = \sqrt{N} [\mu_i - f_\text{th} (X_i)]$.
In this case, $[T^2/(N-1)] [(N-d)/d]$ is distributed as the noncentral
$F$ distribution of $F_{d, N-d}$, which is defined in
Ref.~\cite{anderson-2003}, with noncentrality parameter $\kappa$, defined by
$\kappa = \sum_{i,j} \rho_i \Gamma_{ij}^{-1} \rho_{j} $.
Here, $d$ is the degrees of freedom of the fitting.
In Ref.~\cite{anderson-2003}, it is proved that the limiting
distribution of $T^2$ as $N \rightarrow \infty$ is the
$\chi^2$-distribution with $d$ degrees of freedom if 
$f_\text{th} (X_i) = \mu_i$.

The multivariate statistical theory predicts the following 
\cite{schervish-1995}:
\begin{eqnarray}
{\cal E} (T^2) &=& (d + \kappa) \left[ 1 + \frac{d+1}{N} + 
  {\cal O}(\frac{1}{N^2}) \right] 
  \label{eq:exp_tsq}\\
{\cal V} (T^2) &=& 2 (d + 2\kappa) 
  \biggl[ 1 + \frac{1}{N} 
    \Big(2 d + 4 + \frac{(d+\kappa)^2}{d + 2 \kappa} \Big) 
    + {\cal O}(\frac{1}{N^2} ) \biggl] 
  \label{eq:var_tsq}
\,,
\end{eqnarray}
where ${\cal E} (T^2)$ and ${\cal V} (T^2)$ represent the expectation 
value and variance of the $T^2$, defined in Eq.~\eqref{eq:def_Tsq}.
Here, $d$ is the degrees of freedom and $\kappa$ is the noncentrality
parameter.
If the fitting function is exact (which means $f_\text{th}(X_i) = \mu_i$),
the noncentrality parameter is zero. In that case, if we have large 
enough number of data samples to ignore the $\mathcal{O}(1/N)$ terms,
we expect that the $T^2$ has a value around the degrees of freedom,
$T^2 = d \pm \sqrt{2d}$.

%%%
%%% SUBSECTION: Fitting with inexact fitting function %%%
%%%
\subsection{Inexact fitting function}
One caveat is that the covariance fitting works only if the fitting
function is precise enough.
In practice, we determine the fitting function based on the SChPT
and it is given as a series of ${\cal O}(p^{2n})$.
Since we can include only the finite number of terms in the series,
we usually truncate the series at a certain higher order.
As a consequence, the fitting function has a truncation error which 
makes it inexact in some high precision.
This usually does not cause much trouble.
However, if the covariance matrix has a very small eigenvalue,
$\lambda_l$, the truncation error can be amplified by a factor of
$\dfrac{1}{\lambda_l}$, and then, sometimes, causes failure of the
covariance fitting.

To see this, let us rewrite the Eq.~\eqref{eq:def_Tsq} using the
eigenmode decomposition:
\begin{equation}
\label{eq:eig_mod_decomp}
[C_{ij}^{-1}] = \sum_{k=1}^D \frac{1}{\lambda_k} 
  | v_k \rangle \langle v_k |, 
\qquad
T^2 = \sum_{k=1}^D \frac{1}{\lambda_k}  
  \langle \bar{y} - f_\text{th} | v_k \rangle^2
\,,
\end{equation}
where $\lambda_k$ and $| v_k \rangle$ are eigenvalues and eigenvectors
of the covariance matrix $C_{ij}$, respectively. 
Here, the average data points and 
the fitting function values are also written in bra-ket vector notation, 
$| \bar{y} \rangle_i \equiv \bar{y}_i$ and 
$| f_\text{th} \rangle_i \equiv f_\text{th}(X_i)$.
If an eigenvalue $\lambda_l$ is very small, the $T^2$ is dominated by
the corresponding eigenmode. 
The fitting procedure works very hard to minimize the difference 
between the average data points and the fitting function value, 
$(\bar{y} - f_\text{th})$, in $|v_l\rangle$ direction.
If the fitting function has error in $|v_l\rangle$ direction, the fitting
procedure endeavor to fit in wrong direction, losing precisions
in other directions.
Even if the error of fitting function is small, the lost precisions
in other directions can yield significant error of fitting result.
Section \ref{subsec:bk_cov_fit} exemplifies this situation.

If we have large number of samples, Eq.~\eqref{eq:exp_tsq} and 
Eq.~\eqref{eq:var_tsq} can be approximated by
\begin{equation}
T^2 = d+\kappa \pm \sqrt{2(d+2\kappa)}
\,,
\end{equation}
where $d$ is the degrees of freedom of the fitting and $\kappa$ is 
the noncentrality parameter.
Using the eigenmode decomposition, the $\kappa$ can be written as
\begin{equation}
\kappa = \sum_{k=1}^D \frac{1}{\lambda_k} 
  \langle \mu - f_\text{th} | v_k \rangle^2
\,,
\end{equation}
where $\mu_i$ are the true mean of $\bar{y}_i$.\footnote{Here, we 
assume that we have large enough number of data samples so that 
the $\lambda_k$ and $|v_k\rangle$ of sample covariance matrix 
$C_{ij}$ are fairly representing those of the true covariance matrix.}
Therefore,  the error of fitting function, $(\mu - f_\text{th})$, 
increases the minimized value of $T^2$. Even if the error is small, 
tiny eigenvalues amplify the $\kappa$.
%

%%%
%%% SUBSECTION: Trouble with covariance fitting for $B_K$ %%%
%%%
\subsection{Trouble with covariance fitting for $B_K$}
\label{subsec:bk_cov_fit}
To demonstrate the problem, we choose the $B_K$ data on the C3 
(coarse) ensemble of Ref.~\cite{wlee-10-3}.
This ensemble is particularly a good sample, because it has relatively
large statistics.
It contains 671 configurations and we measured 9 times for each 
configuration.
Details are given in Ref.~\cite{wlee-10-3}.

The fitting functional form suggested by the SU(2) staggered chiral
perturbation theory (SChPT) is linear as follows:
\begin{equation}
f_\text{th} (X) = \sum_{a=1}^{P} c_a F_a(X)
\,,
\label{eq:fit-func-1}
\end{equation}
where $c_a$ are the low energy constants (LECs) and $F_a$ are
functions of $X$, which represents collectively $X_P$ (pion squared
mass of light valence (anti-)quarks), $Y_P$ (pion squared mass of
strange valence (anti-)quarks), and so on.
The details on $F_a$ and $X$ are given in Ref.~\cite{wlee-10-3}.
Here, we focus on the X-fit of 4X3Y-NNLO fitting of the SU(2) SChPT,
which is explained in great detail in Ref.~\cite{wlee-10-3}.
Since we have only 4 data points, we truncated higher order terms
in the fitting function and we have three LECs so $P=3$.
The neglected highest order term in the $f_\text{th} (X)$ is
$X^2 (\ln(X))^2 \approx 0.006$,
where $X = X_P/\Lambda^2 \approx 0.02$.
Hence, the fitting function has an error in that order.

In the X-fit, we fix $am_y = 0.05$ and select 4 data points
of $am_x =$ 0.005, 0.010, 0.015, 0.020 to fit to the functional
form suggested by the SU(2) SChPT as in Ref.~\cite{wlee-10-3}.
Hence, the covariance matrix $C_{ij}$ is a $4 \times 4$ matrix.
Its eigenvalues are
\begin{equation}
\lambda_i 
  = \{\ 1.95\times 10^{-5},\  1.92\times 10^{-6}, 
    \ 7.58\times 10^{-8},\ 1.11\times 10^{-9}
\}
\,.
\end{equation}
Due to the high correlation of data, the smallest eigenvalue is 
smaller than the largest eigenvalue by four orders of magnitude.
Let us look into the eigenvectors,
\begin{equation}
|v_1\rangle = \left[ \begin{array}{r}
   0.837 \\
   0.429 \\
   0.276 \\
   0.200
\end{array} \right] \quad
|v_2\rangle = \left[ \begin{array}{r}
   -0.508 \\
    0.387 \\
    0.542 \\
    0.546
\end{array} \right] \quad
|v_3\rangle = \left[ \begin{array}{r}
    0.202 \\
   -0.739 \\
    0.0725 \\
    0.639
\end{array} \right] \quad
|v_4\rangle = \left[ \begin{array}{r}
    -0.0378 \\
     0.347 \\
    -0.790 \\
     0.503
\end{array} \right]
\,.
\end{equation}
The eigenvector $|v_4\rangle$ corresponds to the smallest eigenvalue
and it dominates the fitting completely.

\begin{figure}[t!]
\centering
\subfigure[Full covariance fit]{\includegraphics[width=0.45\textwidth]
  {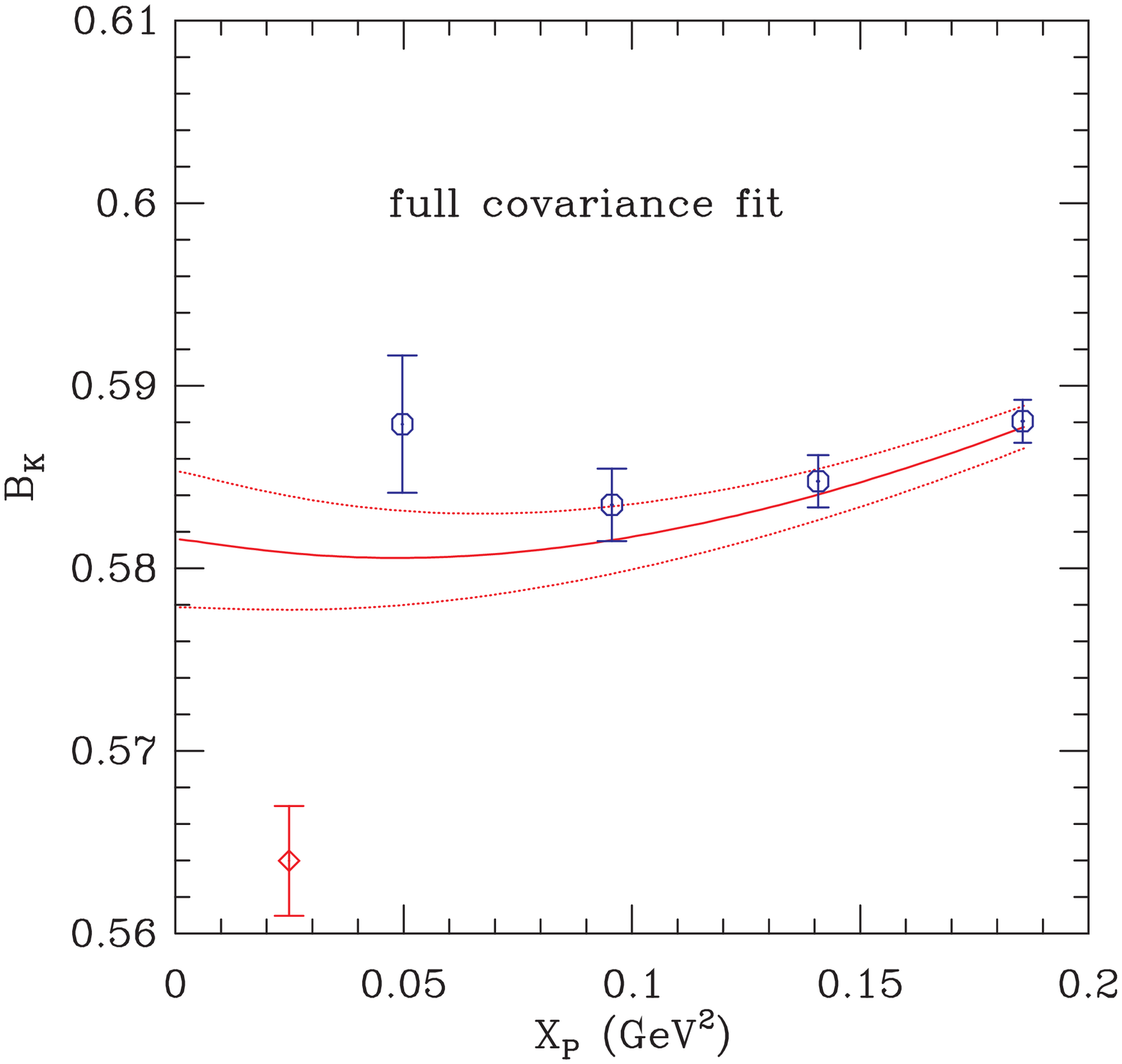}
  \label{subfig:full_cov_fit}}
\subfigure[Diagonal approximation]{\includegraphics[width=0.45\textwidth]
  {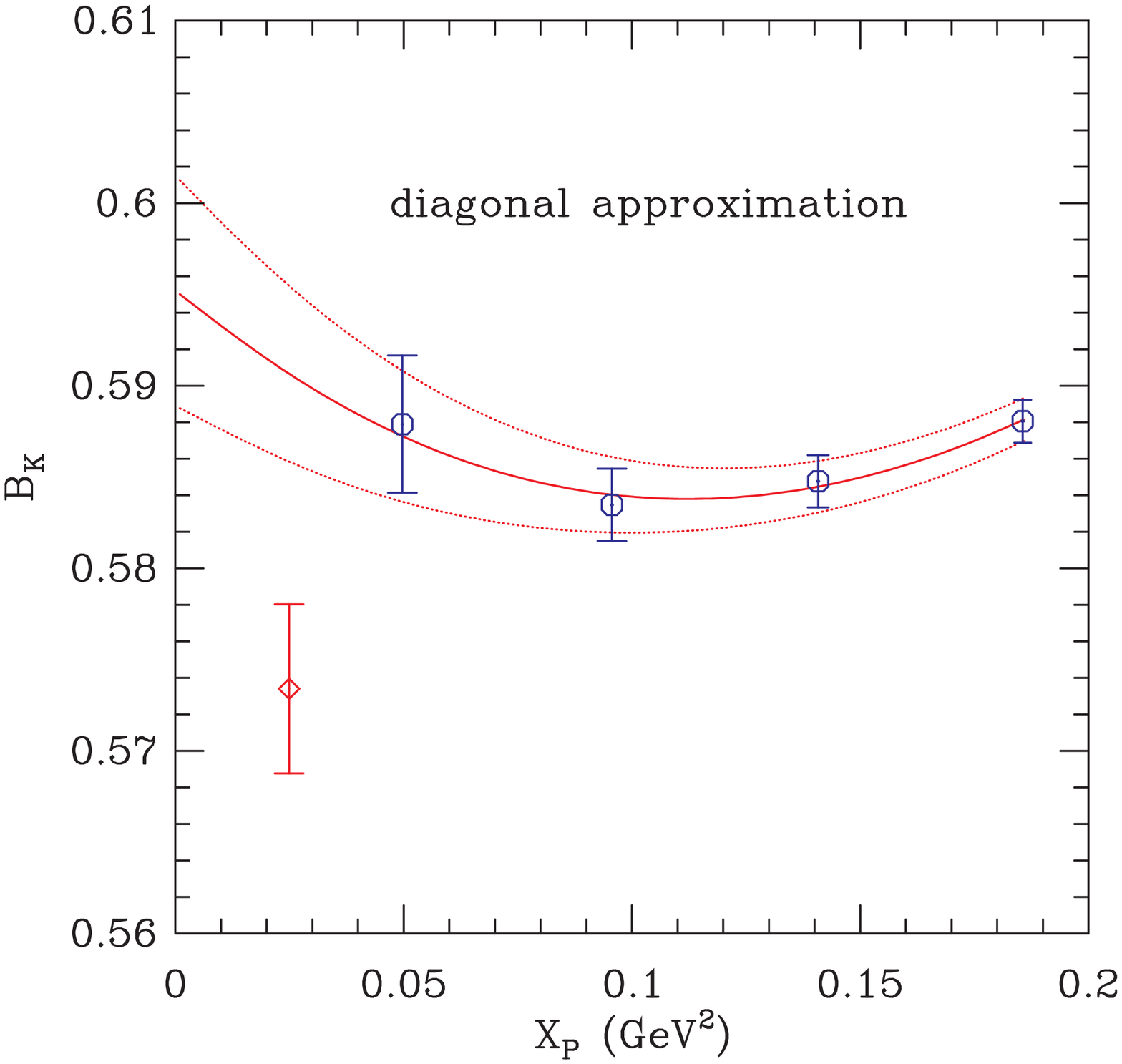}
  \label{subfig:diag_fit}}
\caption{
  $B_K(1/a)$ vs. $X_P$ on the C3 ensemble.  The fit type is
  4X3Y-NNLO in the SU(2) analysis.  The red
  line of left figure represents the results of fitting with the 
  full covariance matrix. The red line of right figure represents 
  the results of fitting with the uncorrelated fitting using 
  diagonal approximation. The red diamond corresponds to the $B_K$ value
  obtained by extrapolating $m_x$ to the physical light valence
  quark mass after setting all the pion multiplet splittings to
  zero.}
\label{fig:full_and_diag}
\end{figure}

In Fig.~\ref{subfig:full_cov_fit}, we show the fitting results with the 
full covariance matrix.
As one can see, the fitting curve does not pass through the data
points.
The $T^2$ value is $7.2$ with degrees of freedom $1$, which indicates
that the fitting fails manifestly.
Let us perform the eigenmode decomposition on $|\bar{y}\rangle$ and
$|f_\text{th}\rangle$ as follows:
\begin{equation}
|\bar{y}\rangle = \sum_{i=1}^4 a_i |v_i\rangle
\,, \qquad
|f_\text{th}\rangle = \sum_{i=1}^4 b_i |v_i\rangle
\end{equation}
where $a_i$ and $b_i$ are the eigenmode projection coefficients.
As we can see in Table \ref{tab:y-f-eigen-cov}, the difference is
$1.75 \sigma$ for $|v_1\rangle$, and $1.7\sigma$ for $|v_2\rangle$,
whereas it is only $0.33\sigma$ for $|v_4\rangle$.
Hence, the procedure of the covariance fitting works hard for the
coefficient of $|v_4\rangle$ but works less precisely for the
coefficients of $|v_1\rangle$ and $|v_2\rangle$, mainly because the
eigenvalue $\lambda_4$ is significantly smaller than $\lambda_1$ and
$\lambda_2$.
The irony is that the average data points, $|\bar{y}\rangle$, has only
$0.015\%$ overlap with $|v_4\rangle$ while more than $99\%$ of them
are dominated by $|v_1\rangle$ and $|v_2\rangle$.
As a result, the fitting function misses the average data points.
In this sense, the failure of the full covariance fitting is obviously
due to the fact that the covariance fitting tries to determine the
coefficient of $|v_4\rangle$ very precisely, while losing precisions
in $|v_1\rangle$ and $|v_2\rangle$ direction.
If the fitting function is exact, this procedure should yield a fitting
result reasonably describing the data.
However, if the fitting function has error in $|v_4\rangle$ direction,
this failing situation can happen.
\begin{table}[htbp]
  \caption{ Eigenmode decomposition of $|\bar{y}\rangle$ and 
    $|f_\text{th}\rangle$ for the full covariance fitting.}
  \label{tab:y-f-eigen-cov}
\begin{center}
\begin{tabular}{c | c c c c }
\hline
\hline
$i$ & 1 & 2 & 3 & 4 \\
\hline
$a_i$     & 1.021(4)   & 0.5655(14) & 0.1061(3)    & 0.01442(3)    \\
$b_i$     & 1.014(4)   & 0.5679(11) & 0.1058(3)    & 0.01443(3)    \\
\hline
\hline
\end{tabular}
\end{center}
\end{table}
%
%
%

%%%%%%%%%%%%%%%%%%%%%%%%%%%%%%%%%%%%%%%%%%%%%%%%%
%%% SECTION :  Prescriptions for the trouble  %%%
%%%%%%%%%%%%%%%%%%%%%%%%%%%%%%%%%%%%%%%%%%%%%%%%%
\section{Prescriptions for the trouble}
If the covariance matrix has small eigenvalues, even a small error of
fitting function may yield large error in fitting result. 
To circumvent this problem, we need some approximation methods, such as
diagonal approximation or cutoff method.
In subsection \ref{subsec:es-method}, we propose a new method which we
call the eigenmode shift (ES) method.

One simple solution to the problem is to use the diagonal
approximation (uncorrelated fitting).
In this method, we neglect the off-diagonal covariance as follows:
$C_{ij} = 0  \text{ if } i \ne j \,.$
In this way, the small eigenvalue problem disappears.
The fitting results are shown in Fig.~\ref{subfig:diag_fit}.

\begin{figure}[t!]
\centering
\subfigure[Cutoff method]{\includegraphics[width=0.45\textwidth]
  {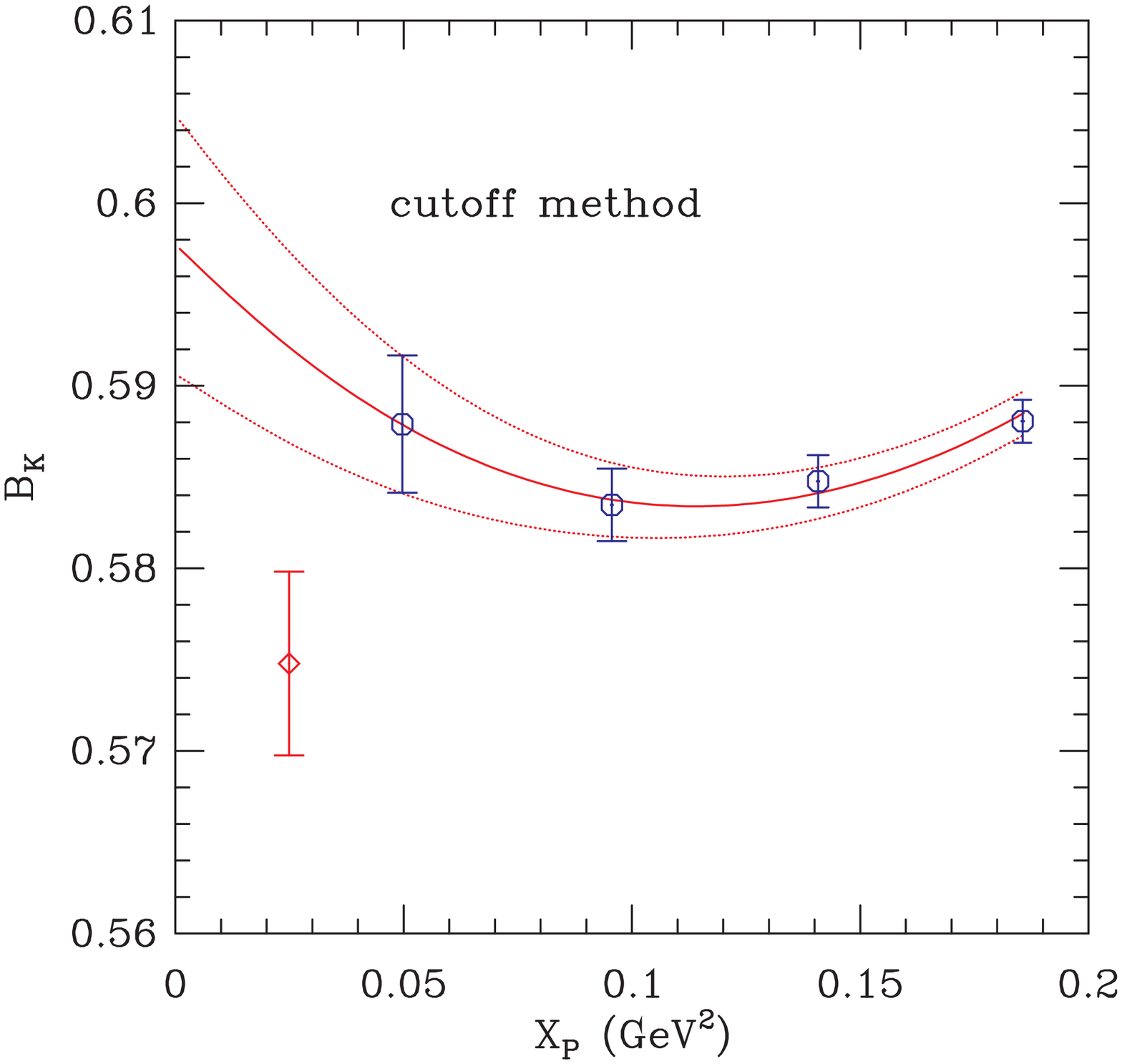}
  \label{subfig:cutoff_fit}}
\subfigure[Eigenmode shift method]{\includegraphics[width=0.45\textwidth]
  {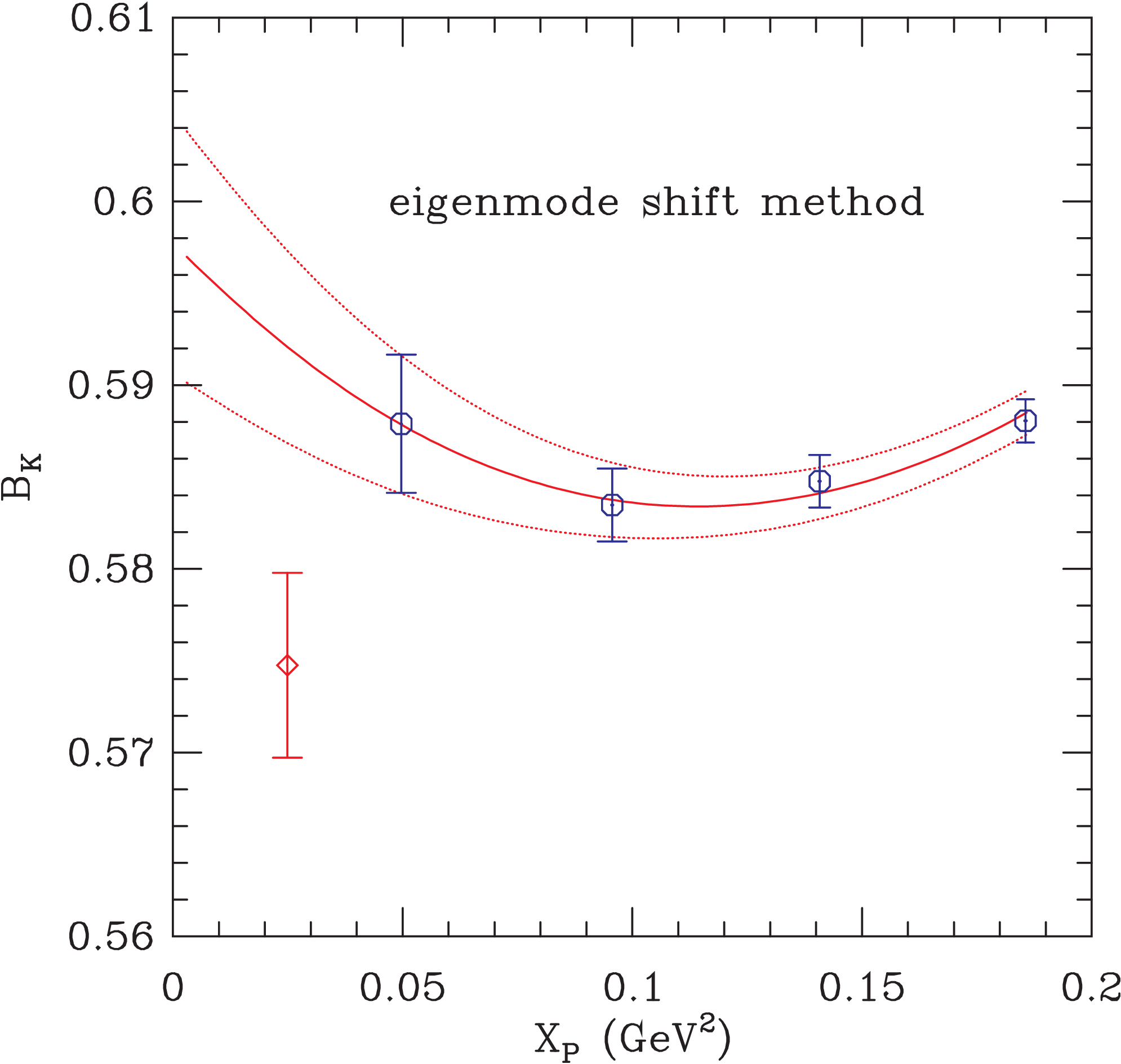}
  \label{subfig:es_fit}}
\caption{
  $B_K(1/a)$ vs. $X_P$ on the C3 ensemble.
  The left figure shows the result of cutoff method and 
  the right figure shows the result of eigenmode shift method.}
\label{fig:cutoff_and_es}
\end{figure}

Another possible solution is to exclude the eigenmodes corresponding to
the small eigenvalues from the inverse covariance matrix, $C^{-1}_{ij}$.
In our example, $|v_4\rangle$ is removed by setting $\dfrac{1}{\lambda_4} = 0$
in Eq.~\eqref{eq:eig_mod_decomp}.
We call this the cutoff method.
A number of lattice QCD groups \cite{fnal-2010-1,lanl-1999-1}
use this method in the popular name 
of the SVD (singular value decomposition) method.
In Fig.~\ref{subfig:cutoff_fit}, we show the results of the covariance
fitting using the cutoff method.

%
%
%

%%%
%%% SUBSECTION: Eigenmode shift method %%%
%%%
\subsection{Eigenmode shift method}
\label{subsec:es-method}

We know that the whole trouble comes from the error of fitting function
in $|v_4\rangle$ direction.
Hence, we can think of a new fitting function $f_\text{th}'$
defined as follows:
\begin{equation}
f_\text{th}'(X) = f_\text{th} (X) + \eta |v_4\rangle
\,.
\end{equation}
Here, $\eta$ is a tiny parameter that can be determined
by the Bayesian method.
Hence, we modify the $\chi^2$ as follows,
\begin{equation}
\chi^2_\text{aug} = \chi^2 + \frac{(\eta-a_\eta)^2}{\sigma_\eta^2}
\,, \qquad
\chi^2 = \sum_{i,j} [\bar{y}_i - f_\text{th}'(X_i)] C^{-1}_{ij} 
     [\bar{y}_j - f_\text{th}'(X_j)]
\,.
\end{equation}
We know that $\eta$ is very tiny so we choose $a_\eta = 0$.
As mentioned in section \ref{subsec:bk_cov_fit}, the order of the 
neglected highest order term in the $f_\text{th} (X)$ is $0.006$.
Hence, we set $\sigma_\eta = 0.006$.
Then we can do the full covariance fitting with an extra fitting 
parameter, $\eta$.
When we do the extrapolation to the physical pion mass,
we use only the $f_\textrm{th}(X)$ function, dropping out the 
$\eta$ terms. 
We call this the eigenmode shift (ES) method.

This is the same procedure as following:
First, find a shifting vector, $\eta |v_4\rangle$, which minimizes the 
$\chi^2_\text{aug}$.
Then fit with the tuned(shifted) fitting function.
To consider the statistical error of $\eta$, do this procedure over
jackknife or bootstrap samples.

In our example, the fitted $\eta = -0.00082(31)$, which is much
smaller by an order of magnitude than truncated highest order terms in
$f_\text{th}$.
In Fig.~\ref{subfig:es_fit}, we show the fitting results obtained
using the ES method.
This method tunes the fitting function by a tiny amount so that
minimizes the small eigenvalue contribution.
In this sense, it looks similar to the cutoff method.
However, unlike the cutoff method, the ES method determines the
shifting parameter, $\eta$, using the Bayesian method and
the full covariance matrix remains untouched.
%

%%%%%%%%%%%%%%%%%%%%%%%%%%%%%%
%%% SECTION :  Conclusion  %%%
%%%%%%%%%%%%%%%%%%%%%%%%%%%%%%
\section{Conclusion}
Here, we address an issue of covariance fitting on the highly
correlated $B_K$ data.
It turns out that the small error of fitting function can make the
fitting fail if the covariance matrix has small eigenvalues.
In order to get around the trouble, we have used approximations: the
diagonal approximation and the cutoff method.
Here, we propose a new method, the eigenmode shift method, which
fine-tunes the fitting function, while keeping the covariance matrix
untouched.

\section{Acknowledgments}
C.~Jung is supported by the US DOE under contract DE-AC02-98CH10886.
The research of W.~Lee is supported by the Creative Research
Initiatives Program (3348-20090015) of the NRF grant funded by the
Korean government (MEST). 
W.~Lee would like to acknowledge the support from KISTI supercomputing
center through the strategic support program for the supercomputing
application research [No. KSC-2011-C3-03].
Computations were carried out in part on QCDOC computing facilities of
the USQCD Collaboration at Brookhaven National Lab, and on the DAVID
GPU clusters at Seoul National University. The USQCD Collaboration are
funded by the Office of Science of the U.S. Department of Energy.

\end{document}